
\magnification=1200
\hyphenpenalty=2000
\tolerance=10000
\hsize 14.5truecm
\hoffset 1.truecm
\openup 5pt
\baselineskip=24truept
\font\titl=cmbx12



\def\rg{R_g}
\def\a{^{'}}
\def\Tg{T_{\gamma}}
\def\({\left(}
\def\){\right)}
\def\rhoz{\rho_{0}}

\def\kes{\kappa_{es}}
\def\kff{\kappa_{ff}}
\def\ref{\par\noindent\hangindent 20pt}
\def\refig{\par\noindent\hangindent 15pt}
\def\mincir{\raise -2.truept\hbox{\rlap{\hbox{$\sim$}}\raise5.truept
\hbox{$<$}\ }}
\def\magcir{\raise -4.truept\hbox{\rlap{\hbox{$\sim$}}\raise5.truept
\hbox{$>$}\ }}
\def\rho{\varrho}
\def\Mdot{\dot M}
\def\mdot{\dot m}

\def\Menv{M_{env}}
\null
\vskip 1.2truecm
\centerline{\titl WINDS FROM NEUTRON STARS}
\centerline{\titl AND STRONG TYPE I X--RAY BURSTS}
\vskip 1.5truecm
\centerline{Luciano Nobili, Roberto Turolla}
\medskip
\centerline{Department of Physics, University of Padova}
\centerline{Via Marzolo 8, 35131 Padova, Italy}
\vskip 0.3truecm
\centerline{and}
\vskip 0.3truecm
\centerline{Iosif Lapidus \footnote{$^{\ddag}$}{The Royal Astronomical Society
Sir Norman Lockyer Fellow}}
\medskip
\centerline{Institute of Astronomy, University of Cambridge}
\centerline{Madingley Road, Cambridge CB3 0HA, UK}
\vfill\eject

\beginsection ABSTRACT

A model for stationary, radiatively driven winds from X--ray bursting neutron
stars is presented. General relativistic hydrodynamical and radiative
transfer equations are integrated from the neutron star surface outwards,
taking into account for helium nuclear burning in the inner, dense, nearly
hydrostatic shells. Radiative processes include both bremsstrahlung
emission--absorption and Compton scattering; only the frequency--integrated
transport is considered here. It is shown that each
solution is characterized by just one parameter: the mass loss rate $\Mdot$,
or, equivalently, the envelope mass $\Menv$.
We found that, owing to the effects of Comptonization, steady, supersonic
winds can exist only for $\Mdot$ larger than a limiting value $\Mdot_{min}
\approx\Mdot_{E}$. Several models, covering about two decades
in mass loss rate, have been computed for given neutron star parameters. We
discuss how the sequence of our solutions with decreasing $\Menv$ can be used
to follow the time evolution of a strong X--ray burst during the
expansion/contraction
phase near to the luminosity maximum. The comparison between our numerical
results and the observational data of Haberl {\it et al.\/} (1987) for the
bursts in 4U/MXB 1820-30 gives an estimate for both the spectral
hardening factor and the accretion rate in this source.

\bigskip\bigskip
\noindent
{\it Subject headings:\/} hydrodynamics \ -- \ radiative transfer \ -- \
stars: individual (4U/MXB 1820-30) \ -- \ stars: neutron \ -- \
stars: winds \ -- \ X--rays: bursts

\bigskip\bigskip
\centerline{Accepted for publication in the Astrophysical Journal}
\vfill\eject

\beginsection I. INTRODUCTION

Mass loss from stars is a well--known phenomenon and various models have
been proposed to explain it in different situations. In the case of winds
from hot stars, radiation pressure in lines is often
assumed to be responsible for the outflow while for cooler stars
radiation pressure on dust grains may be the dominating mechanism (see e.g.
the review by Cassinelli 1979). A general feature of all steady--state models
is that the flow is subsonic at small radii and supersonic far from the star,
passing through a critical point which, however, not necessary coincides with
the sonic point.

As far as winds from neutron stars are concerned, several mechanisms
were considered in the attempt to model different phaenomena.
Winds from young and very hot neutron stars, for example, can be driven
by absorption of high--energy neutrinos by protons and neutrons close to the
stellar surface (Salpeter \& Shapiro 1981; Duncan, Shapiro \& Wasserman
1986). Under less extreme conditions, however, the outflow is commonly thought
to be powered by radiation pressure.
In the optically thick models of Meier (1982a, b, c) mass and energy
are injected at some radius and  the rates of injection are free parameters.
For high energy input rates (well above the Eddington limit) most of the energy
is converted into gas kinetic energy and a strong wind is produced, although
velocities are never relativistic.

Highly relativistic winds were considered by Paczy\'nski (1986),
Goodman (1986) and Paczy\'nski (1990) as models for $\gamma$--ray bursters at
cosmological distances. The idea is that $\gamma$--ray bursts may result
from the merging of the two components in a double neutron star
binary. As in Meier's models, energy and mass
are injected at some arbitrary rate near the star surface and
local thermodynamic equilibrium (LTE) is assumed to hold.
Owing to the very high temperature, electron--positron pairs are formed and
the pair pressure is taken into account along with the radiation pressure.
The apparent temperature of the photosphere is in excess of $10^{9}$ K and
the object exhibits itself as a strong $\gamma$--ray source.

The existence of winds from neutron stars during strong X--ray bursts
is a well acknowledged fact (see e.g. Lewin, Vacca \&
Basinska 1984; Tawara {\it et al.\/} 1984; Tawara, Hayakawa \& Kii 1984;
Haberl {\it et al.\/} 1987; see also Lewin, Van Paradijs \& Taam 1993 for
a very
recent review on X--ray bursters). For a distant observer the wind phase lasts
just the few seconds separating the precursor from the main burst (Haberl
{\it et al.\/}). Since the true beginning of the burst is marked by the
precursor, the dip observed
in the fitting temperature curve is an observational evidence of a
photospheric expansion produced, most probably,  by a super--Eddington
luminosity.

Steady--state winds from neutron stars were the object of several
investigations
in the past decade. The possible existence of such winds was explained by
the fact that, during thermonuclear helium burning, the temperature at the
base of the accreted atmosphere rises above $10^{9}$ K, so that the electron
scattering opacity decreases below its Thomson value owing to Klein--Nishina
corrections. The radiation flux diffusing out of the hot region may be
therefore below the local Eddington value, but it may appear to exceed the
local Eddington limit in the outer layers which are cooler. The
hydrostatic equilibrium of these layers is then violated and a
radiation--driven wind may develop.

Earlier studies (Ebisuzaki, Hanawa, \& Sugimoto 1983; Kato 1983;
Quinn \& Paczy\'nski 1985) assumed Newtonian gravity and dealt
only with optically
thick outflows. General relativistic effects were taken into account by
Paczy\'nski and Pr\'oszy\'nski (1986), maintaining the diffusion
approximation, and by Turolla, Nobili \& Calvani (1986), who investigated also
the effects of Compton heating--cooling in the wind. A more refined treatment
of radiative transfer in the outflowing envelope was introduced by Joss \&
Melia (1987), accounting also for Compton scattering in an approximate way.
Using the wind structure
computed by Ebisuzaki {\it et al.\/} as a background for
frequency--dependent radiative transfer, Lapidus (1991) has confirmed the
qualitative scenario of drastic spectral softening during the photospheric
expansion, finding a satisfactory agreement between the computed spectral
softening factors and those observed by EXOSAT in 4U/MXB 1820--30.
In a recent paper Titarchuk (1993) presented an analytical study of
spectral formation during expansion and contraction phases of X--ray
bursters and found that the spectrum in the expansion phase depends strongly
on the temperature in the helium burning region.

The models of neutron star winds explored up to now contain, however, some
drastic assumptions.
Nearly all previous investigations, in fact, neglected the effects of dynamics
on radiative transfer and used LTE plus diffusion approximation everywhere
in the flow. As a result, these
models imply a discontinuity in the angular distribution of the
radiation field which switches from near--isotropy below the photospheric
radius to radial streaming above it, where an optically thin region with
constant outflow velocity is assumed to exist. Moreover the rate of energy
injection at the base of the flow was assumed to be a free parameter,
uncorrelated to the energy actually released by thermonuclear reactions during
the burst phase.

In this paper we present a wind model which overcomes these major
disadvantages. General relativistic radiative transfer in a differentially
moving medium is properly treated using Thorne's (1980) moment formalism, so
that not only the effects of thermal but also of dynamical
Comptonization are accounted for. Integration of the flow equations is pushed
down to the stellar surface and covers also the dense, inner shells in which
nuclear burnings occur. As a consequence, the energy input for our solutions
is self--consistently computed and the only free parameter is the total mass
of the envelope $M_{env}$. We have found that the inclusion of Comptonization
results in a lower limit for $\Mdot$ (and, correspondingly for $M_{env}$)
for steady, supersonic winds to exist, the lower bound being $\Mdot\sim 6
\times 10^{17} \ {\rm g/s}$ for a neutron star mass of $1.5 M_\odot$.
Several models with different
values of $M_{env}$, corresponding to mass loss rates in the range $10^{17}\div
10^{19}$ g/s were computed. Our series of models with smoothly
varying $M_{env}$ can be used to follow the time evolution of a
X--ray bursting source during the envelope expansion/contraction phase. A
comparison of
our results with the observational data of 4U/MXB 1820-30 allowed us to
estimate the hardening ratio and the initial envelope mass for the bursts
analyzed by Haberl {\it et al.\/}

Further developments will be aimed to include frequency--dependent radiative
transfer in our code to obtain a self--consistent determination of
the emergent spectrum. Frequency--dependent calculations are needed also
to access the effect of bulk motion Comptonization,
which proved to be relevant in near--Eddington accretion
onto neutron stars (Zampieri, Turolla \& Treves 1993), and to provide a
better determination of spectrum--averaged
quantities which enter the dynamical equations, like the opacity
coefficients and the radiation temperature. This work is now
in progress and, when completed, the scheme will provide a tool for unambiguous
determination of mass--radius relation for neutron stars from observational
spectral data of powerful X--ray bursts.

\beginsection II. THE MODEL

In the following we assume that the gas outflow is spherically symmetric and
stationary. The neutron star rotation is neglected so that the gravitational
field can be described by the vacuum Schwarzschild solution;
$M_*$ and $R_*$ will denote the star mass and radius, respectively.
We also ignore all the effects induced by the large--scale {\bf B}--field
associated with the neutron star. These hypotheses should be satisfied
for X--rays bursters, which are commonly thought to contain old neutron stars,
and were the starting point of previous investigations on this subject
(see e.g. Kato 1983; Quinn \& Paczy\'nski 1985; Paczy\'nski \&
Anderson 1986; Paczy\'nski \& Pr\'oszy\'nski 1986; Paczy\'nski 1990). All these
studies dealt with radiative wind acceleration in an optically thick
plasma and assumed that the only contribution to opacity comes from
electron scattering. Moreover the rate of energy release by thermonuclear
reactions at the base of the envelope, $\dot E$, was not computed
self--consistently but, together with the mass loss rate $\Mdot$, is a free
parameter of the model. In this investigation we present neutron star wind
models which overcome these limitations and account properly for both energy
production by nuclear burnings and radiative transfer outside the diffusion
regime. The capability of the model to handle radiative transfer under general
conditions
is indeed crucial since, contrary to what is often assumed, a large value of
the scattering depth is not enough, by itself, to guarantee that LTE is
established. This will also allow for a more natural specification of
boundary conditions, which can be placed at radial infinity where the optical
depth vanishes,
avoiding the need to impose ``artificial'' conditions at the photosphere,
as in Paczy\'nski \& Pr\'oszy\'nski (1986) and Paczy\'nski (1990).

The set of equations governing the dynamics of the
gas and the transfer of radiation in spherical, stationary flows in a
Schwarzschild gravitational field are discussed in the paper by
Nobili, Turolla \& Zampieri (1991, hereafter NTZ) to which we refer for all
details.
Although explicitly written for spherical inflows, their equations hold the
same for winds, just reversing the sign of the flow velocity and changing
$\Mdot$ into $-\Mdot$. They are

$$ \eqalignno{
 & w_1^{'} + v w_0^{'} + v w_2 \( {{u}^{'}\over {u}} - 1 \)
   + 2 w_1 \( 1 + {y^{'}\over y} \)
   + {4\over 3} v w_0 \( {{u}^{'}\over {u}} + 2 \) =\cr
 & = {{rR_g}\over y}\left( s_0 + {{\rhoz\epsilon_{nuc}}\over c}\right) & (1)\cr
 & w_2^{'} + v w_1^{'} + {1\over 3} w_0^{'}
   + w_2 \(3 + {y^{'}\over y} \)
   + 2vw_1 \( {{u}^{'}\over {u}} + 1 \)
   + {4\over 3} {y^{'}\over y} w_0 =\cr
 & ={{rR_gs_1}\over y}\  & (2)\cr
 & (v^2-v_s^2){{u\a}\over{u}}-2v_s^2+{1\over{2y^2r}} -
   {{r\rg}\over{u(P+\rho)}}[(\Gamma -1) s_0 - vs_1] = 0 & (3)\cr
 & {{T\a}\over T} - (\Gamma -1){{\rhoz\a}\over\rhoz} +
   {{r\rg s_0}\over{Bu(P+\rho)}}=0 & (4)\cr
 & {{u\a}\over{u}} + {{\rhoz\a}\over\rhoz} + 2 =0\, .
             & (5)\cr}$$
Here and in the following $r = R/\rg$ is the adimensional radial coordinate in
units of the gravitational radius,
a prime denotes derivation with respect to $\ln r$, $v$ is the
fluid velocity measured by a stationary observer in units of $c$,
$y = \sqrt{(1-1/r)/(1-v^2)}$ and $u\equiv yv$; all the other symbols have their
usual meaning. The variable Eddington factor is assumed
to be a given function of the scattering depth $\tau$
$$f_E = {{w_2}\over{w_0}} = {2\over 3}{1\over{1+\tau^n}};$$
$n=2$ was used in the numerical calculations. This approximation for the
closure should guarantee that the radiation
field is correctly computed up to $\sim 15 \%$ (see Turolla \& Nobili 1988).
We have assumed the plasma to be a perfect, fully ionized  gas: $X$, $Y$
and $Z$ denote the mass abundances of hydrogen, helium and metals. The chemical
composition is taken to be constant through the envelope (see the discussion
in section III).
The present set of equations differ slightly from those of NTZ inasmuch as
we included
also the rate of energy generation by nuclear reactions, $\epsilon_{nuc}$.
In order to keep our treatment simple we just take into account 3--$\alpha$
He--burning, although actual nuclear reaction networks in X--ray bursting
neutron stars are much more complicated (see e.g. Taam 1985 for a review).
The expression for the energy generation rate is then (see e.g. Clayton 1968)
$$\epsilon_{nuc} = 3.9\times 10^{11} f(1 - X)^3\rhoz^2 T_8^{-3}\exp\left(
-{42.94\over{T_8}}\right) \ {\rm erg\,g^{-1}s^{-1}}\, ,$$
$f\simeq \exp{[2.76\times 10^{-3}\rhoz^{1/2} T_8^{-3/2}]}\, .$
The source moments $s_0$ and $s_1$ account for both free--free
emission--absorption and Compton scattering. They can be derived from the
expressions given in NTZ, replacing the complete hydrogen cooling function
with bremsstrahlung emissivity
$$s_0 = \kes\rhoz w_0\left[{{\kff^P}\over{\kes}}\left({{aT^4}\over{w_0}} -
1\right) + 4{{kT}\over{m_ec^2}}\left(1 - {{\Tg}\over T}\right)\right]$$
$$s_1 = -\kappa^R\rhoz  w_1\, ,$$
where $\kff^P$ and $\kappa^R$ are the free--free Planck and Rosseland mean
opacities. The radial dependence of the radiation
temperature $\Tg$, needed to evaluate the Compton heating--cooling term in
$s_0$, is obtained from the equation
$${{\Tg\a}\over{\Tg}} = Y\, \left(1-{{\Tg}\over T}\right)\eqno(6)$$
where $Y=(4kT/m_ec^2)\max (\tau,\tau^2)$ is the Compton parameter
(Park \& Ostriker 1989, NTZ).

Equations (1)--(6) can be solved to get the run of all physical quantities
characterizing the wind once boundary conditions and a set of constituent
relations are given. Numerical problems, however, are going to arise in the
dense wind region close to the star surface where the scattering depth,
$\tau$,
becomes very large. Equation (6) shows, in fact, that $\Tg$ must be extremely
close to $T$ for $\tau\gg 1$ and, if $\tau$ is so large to produce an
effective depth, $\tau_{eff} = [3\tau_{ff}(\tau +\tau_{ff})]^{1/2}$,
larger than unity, also the radiation energy density
will nearly equal $aT^4$. Under such conditions the differences
$(1 - \Tg/T)$ and $(1 - aT^4/w_0)$ can become dangerously close to machine
precision, producing unbound errors. To avoid this possibility we decided
to split the integration range into two parts: an outer region
$r_{fit}< r < r_\infty$ (region I), where equations (1)--(6) can be safely
used, and an inner region $r_*< r < r_{fit}$ (region II), where we adopted
the diffusion limit. The value
of the fitting radius $r_{fit}$, where the two branches of the solution must
be matched, is arbitrary but it must be chosen in such a way that equations
(1)--(6) can be integrated without troubles and, at the same time, $\tau_{eff}
(r_{fit})\gg 1$ for diffusion and LTE to hold. The diffusion limit of equations
(1)--(4) is readily obtained by imposing $w_0 = aT^4$ (see Flammang 1982,
Turolla {\it et al.\/}) and yields

$$\eqalignno{
 &   {{(yT)\a}\over{(yT)}} + {\ell\over{8\alpha v_c^2yr}}= 0 & (7)\cr
 &   (v^2 - v_c^2){{u\a}\over u} + v_c^2\left({{T\a}\over T} - 2\right)
     + {1\over{2y^2r}}\left(1 - {{\kappa^R}\over{\kes}}{{y\ell}\over h}
     \right)= 0 & (8)\cr
 &   \ell\a + 2 {{y\a}\over y}\ell + {{\mdot v_c^2}\over y}\left[
     \left(12\alpha + {3\over 2}\right){{T\a}\over T} - (1+4\alpha )
     {{\rhoz\a}\over{\rhoz}}\right] - {{2\kes\rg^2\rhoz r^3}\over{yc^3}}
     \epsilon_{nuc}= 0  & (9)\cr
  &  2\kes\rg\rhoz yvr^2 = \mdot\, .& (10)\cr}$$
Here $\ell$ is the luminosity in units of $L_E = 4\pi GMc/\kes$, $\mdot =
\Mdot c^2/L_E$
is the mass loss rate in units of the critical rate, $\alpha =
P_{rad}/P_{gas}$ and $v_c$ is the isothermal sound speed. Some terms which
become important only for $v\sim 1$ were dropped in equations (7)--(9).

Numerical models show that temperature never exceeds $\sim 10^7$ K
at the
fitting radius so that no relativistic corrections both to the equations
of state and to the opacities are needed in region I.
On the other hand,
since temperature and density at the base of the envelope
must be high enough to make nuclear burning effective, $T(R_*)\sim 4\times 10^9
$ K, $\rhoz(R_*)\sim 10^6 \ {\rm g/cm}^3$ in the case of 3--$\alpha$ reactions,
electrons will be partially degenerate and relativistic in the deep layers.
The form of the equations of state used in the numerical code was obtained
by least--square fitting to Chandrasekhar's (1939) tables the two functions
$$\eqalign{
&  q_1(\ln\rhoz,\ln T) = \ln{P\over{\rhoz T}} \cr
& q_2(\ln\rhoz,\ln T) = \ln{{2U}\over{3P}} \cr}$$
where $U = \rho - \rhoz c^2$ and $q_1$, $q_2$ are double Chebyshev
series. Truncating the sums to sixth degree gives an accuracy $\mincir 5$ \% .
The assumption of neglecting electron conduction appears to be
justified in our case. The ``effective'' opacity $\kappa$, including
both conductive $\kappa_c$ and radiative contributions, has the form (see
e.g. Clayton) $1/\kappa = 1/\kappa_c + 1/\kes$. In the burning region,
temperature and density are such that $\kappa_c > \kes$ and this
inequality becomes even stronger moving to larger radii.
Klein--Nishina corrections to the scattering opacity were included, using
Paczy\'nski (1983) interpolating formula. We checked the validity of this
expression by computing the Rosseland mean of the frequency--dependent
Compton transport opacity of Shestakov, Kershaw \& Prasad (1988).
Agreement is within 10 \% for $T\mincir 4\times 10^9$ K.

The actual form of the boundary conditions for the two sets
of equations will be discussed in detail in the next section. Here we want to
analyze how boundary conditions should be placed for
equations (1)--(6) if they were integrated in the whole range $r_*< r <
r_\infty$, since this will make apparent the number of free parameters of our
model. At radial infinity radiation must stream freely, hence we have to
require that
$$w_0 = w_1\qquad\qquad\qquad {\rm at} \ r=r_\infty\, .$$
On the neutron star surface we assume that LTE is established and that
the radiative flux goes to zero since all luminosity is generated by nuclear
reactions above $r_*$. This amounts to ask that
$$\eqalign{&w_0 = aT^4  \cr
           &\Tg = T     \qquad\qquad\qquad{\rm at} \ r=r_*\cr
           &w_1 = 0\, .    \cr}$$
A further condition has to be imposed at the sonic point where $v = v_s$ to
ensure the regularity of the solution, thus leaving just one degree of freedom
which can be the value of either velocity, density or temperature at a given
radius, or, alternatively, the mass loss rate $\Mdot$.

\beginsection III. RESULTS AND DISCUSSION

In this section we present the results of the numerical integration of the
equations of radiative hydrodynamics for the wind case. All our models refer
to a neutron star of mass $M_* = 1.5 M_\odot$ and radius $R_* = 3 \rg =
13.5$ km; the chemical composition of the outflowing material can be varied
to explore its effects on the wind properties. The numerical code makes
use of a relaxation technique (Nobili \& Turolla 1988) and the integration
range extends from $r=3$ to $r=10^5$; the upper limit is fixed essentially
by the requirement that the scattering depth become low enough to make the
radial streaming approximation reasonable. As we discussed in the previous
section, the integration domain is split into two parts and equations
(1)--(6) are used for $r> r_{fit}$ (region I) while the diffusion limit,
equations (7)--(10), is assumed for $r< r_{fit}$ (region II); the complete
solution is then obtained by fitting the two branches. The procedure is
the following. First $r_{fit}$ is fixed and equations (1)--(6) are integrated
for $r> r_{fit}$ with the conditions $T = T_{fit}$, $\Tg = T$, $w_0 =
aT^4$, $v= \epsilon v_s$ at $r_{fit}$ and $w_0 = w_1$ at radial infinity; here
$\epsilon\leq 1$ is a parameter and
boundary conditions must be supplemented with the regularity condition at the
sonic point
where $v = v_s$. Once the solution in region I is known, equations (7)--(10)
are solved in region II, subject to the conditions that $\mdot$, the velocity
and the velocity gradient at $r_{fit}$ are those provided by the solution we
have just
computed and that $L$ vanishes at $r = r_*$. In this way all the variables and
their derivatives are continuous at $r_{fit}$, with the exception of
temperature. Finally, the continuity of $T$ is achieved by fine--tuning the
parameter $\epsilon$ so that the final model is characterized only by the
value of $T_{fit}$ or, equivalently, by the mass--loss rate $\mdot$.
In practice we found numerically more convenient to keep $T_{fit}$ fixed,
$T_{fit} = 2.5\times 10^7$ K, and to vary $r_{fit}$; all solutions have
$25<r_{fit}<150$. Since in our numerical code the exact form of the
critical point condition is not so crucial (see the discussion in NTZ),
we just asked that $(yL)\a = 0$ at $v = v_s$.

A sequence of models was obtained, covering nearly two decades in mass loss
rate from $\mdot\sim 1$ up to $\mdot\sim 100$. Chemical composition
was assumed to be nearly solar, with mass abundances $X=0.6$,
$Y=0.35$, and $Z=0.05$. The radial dependence of some
physically significant quantities is presented in figures 1--5
for three characteristic
values of $\mdot$, namely: a) $\mdot = 2.8$, b) $\mdot = 50.2$, and
c) $\mdot = 102.3$. In figures 1a, b, c
the run of bulk velocity, sound speed and density is presented;
the crossing of the two velocity curves marks the sonic point. Terminal
wind velocities
are never relativistic and do not exceed $\sim 3\times 10^{-3}\, c
\sim 1000$ km/s. Our values for $v_\infty$ are systematically lower
than those obtained by Paczy\'nski \&
Pr\'oszy\'nski, as it should be expected,
because of the stronger coupling between matter and radiation when
diffusion is assumed, and also of those presented by Joss \& Melia.

Figures 2a--c show the radial distribution of the gas and radiation
temperatures, $T$ and $\Tg$. The filled dots mark the
points where the matching between the inner (diffusive) and outer regions
was achieved, in order to illustrate the smoothness of the numerical fitting.
At the point where the curves of $T$ and $\Tg$ start to
diverge, LTE between radiation and matter breaks down;
at larger $r$, $\Tg$ stays constant since matter and radiation
are decoupled. The behavior of $T$ reminds
qualitatively that one of steady atmospheres of X--ray bursters in
the contraction phase (Lapidus, Sunyaev, \& Titarchuk 1986; London, Taam, \&
Howard 1986) and is similar to that found by Joss \& Melia, although their
treatment of Comptonization is different from the present one. The decrease
of $T$ with $r$ after matter has decoupled
from radiation is halted by the Compton heating of the cooler electrons by the
photons originating in the
inner, much hotter layers. Such an effect is much more pronounced for low
$\mdot$ models because they exhibit a sufficiently large translucent
($\tau_{eff}<1$ and $\tau >1$) region, contrary to high $\mdot$ ones.
At even larger radii adiabatic cooling, due to $PdV$
expansion, becomes more efficient than Compton heating, and $T$ decreases
again. We stress that, contrary to previous investigations, the temperature
at the base of the flow, $T_b$, is now self--consistently determined, since we
have taken into account nuclear burnings in the expanding envelope.
3--$\alpha$ reactions actually keep $T_b\sim 3\times 10^9$ K for all
values of the mass loss rate, because of the strong temperature dependence
of the reaction rate.

The radial run of luminosity, as measured by the comoving observer in
Eddington units, is given in figure 3 for all three models.
Luminosity rises from zero at $R=R_*$, reaches its maximum extremely close
to the stellar surface, and then decreases at larger $r$ as radiative flux
is converted into bulk kinetic energy. The larger $\mdot$ is, the higher
the peak luminosity is, to cope with the larger flow inertia. Nuclear
reactions can always produce the luminosity required to propel the wind
because larger envelope masses result in higher densities which, in turn,
enhance the nuclear reaction rate. Luminosity at infinity is always extremely
close to the Eddington value, as indeed should be, since all the exceeding
power is transferred to the matter flow.

Figures 4 and 5 illustrate the radial runs of optical depths and
radiation moments just for model b), the overall behaviour being similar for
the other models; for the sake of clarity only the outer region is presented.
In figure 4, together with the scattering depth, $\tau$ and
the effective depth, $\tau_{eff}$,
the product $\tau v$ is also shown since this parameter gives
a measure of bulk motion Comptonization (Payne \& Blandford 1981; Nobili,
Turolla \& Zampieri 1993). Bulk motion Comptonization is, however, expected
to be efficient only in regions where $\tau_{eff}\mincir 1$, $\tau
\magcir 1$ and $\tau v\magcir 0.1$; as can be seen from the graph,
$\tau v\mincir 0.1$ above the thermalization radius and dynamical
effects are never dominant.
The transition between the diffusive and the streaming regime is clearly
visible in figure 5: at large optical depth $w_0\sim\tau w_1$ while,
above the last scattering radius, $w_0\simeq w_1$.

Although, as we already pointed
out, each solution is characterized by the value of $\Mdot$, it is much more
physically meaningful to label models with the total mass contained in the
``static'' part of the envelope. We define $\Menv$ as the mass contained in
between the base of the nuclear burning shell (which is assumed to coincide
with the neutron star surface) and the sonic radius, $R_s$,
$$\Menv = 4\pi\int_{R_*}^{R_s}\, dR\, R^2\rhoz\, .$$
Calling this portion
of the atmosphere ``static'' seems reasonable, since below the sonic radius
dynamics does not produce major changes in the structure with respect
to the hydrostatic case. The envelope mass as a function of
$\mdot$ is shown in figure 6 for all computed models.
The importance of $\Menv$ is that, contrary to
$\Mdot$, it can be used to characterize both the pre--burst phase and the
time evolution during the photospheric expansion/contraction phase. In fact,
the value
of the envelope mass when the expansion begins is proportional,
assuming a constant mass transfer rate from the companion star, to the
time between two successive bursts.
Furthermore, the temporal evolution of a single strong burst with expansion,
at least close to the luminosity peak, can be thought as a sequence of
quasi--stationary models with gradually decreasing $\Menv$.
This decrease is mostly
due to the fact that the nuclear burning shell moves outwards leaving the
products behind (and out of the wind region); there is, in addition, a
small decrease of $\Menv$ in time because some mass is actually lost
through the wind itself. The fact that radiative luminosity pushes the
material only above the He--burning shell together with the thinness of the
shell makes the constant composition assumption reasonable, although a
composition gradient will be present across the burning region. We do not
expect this to modify our results significantly since the variation in the
chemical composition will affect, at most, the inner two or three radial
zones.

The fact that,
in order to start the wind, a sufficient amount of material should be
accumulated onto the neutron star surface is unanimously accepted. The
expansion phase, however,
according to observational data, lasts just $\mincir 10$ seconds, so one has
to face the problem of quenching the wind in quite a short time. It is
usually assumed
that the wind ceases when nearly all the nuclear fuel is exhausted and the
comoving luminosity at the base of the envelope drops below a critical value,
which can be derived from the energy conservation.
The characteristic time scale of the process is
then $t_{nuc} = \epsilon Y\Menv c^2/\dot E$, where $\epsilon=6.1\times
10^{-4}$ is the efficiency of 3--$\alpha$ reactions (see e.g. Clayton
1968) and $\dot E\sim (1 + \mdot)L_{E}$ is the total (radiative plus
advected plus kinetic) luminosity.
Clearly, no lower limit for the mass loss rate
follows from energetic considerations, and winds with arbitrarily low $\mdot$
are possible.
On the contrary, we have found that a lower
limit for the mass loss rate definitely
exists, and the presence of such a bound is due to Compton heating. This
effect is analogous to the so called ``preheating'' limit in spherical
accretion (see e.g. NTZ). As we already discussed, Compton heating is
stronger for
low $\mdot$ and tends to isothermize the outflow at $T\sim 10^7$ K, inhibiting
the decrease of sound speed with radius: as a consequence there may be
problems for the flow to become supersonic, since the sonic point will
move to larger and larger radii. To see this in more detail
let us consider the momentum
equation [eq. (3)]; since the velocity gradient must be positive in the
subsonic region, it follows that

$$2v_s^2 - {1\over{2r}} +{{r\rg}\over{u(P+\rho )}}\left[(\Gamma -1)s_0 -
vs_1\right]<0\, .\eqno(11)$$
Condition (11) can be written in a more transparent form
using the relation $w_1 = c^2\ell/(2\kes r^2\rg)$ and taking into account
that above $r_{fit}$, where Comptonization is effective, true
emission--absorption can be neglected, $P+\rho\sim\rhoz c^2$ and $y\sim 1$
$$2v_s^2 + {1\over{2r}}(\ell -1 ) + {2\over 3}{{\kes\rhoz r\rg}\over{\mdot}}
{{w_0}\over{w_1}}{{4kT}\over{m_ec^2}}\left(1 - {{\Tg}\over{T}}\right)<0\, .
\eqno(12)$$
The physical meaning of the various terms in this expression is
straightforward: the first term accounts for the gas pressure force, the
second one represents the effective radiative force while the last one
is the first order correction, due to non--conservative scatterings, to the
Thomson radiative force exerted by the outgoing radiation on electrons.
The Compton correction can be either positive or negative, according to
the value of $\Tg/T$. For $\Tg <T$ it acts like an extra thrust, and
gives rise to the so called Compton rocket (O'Dell 1981; Cheng \& O'Dell
1981). Under our conditions $\Tg >T$, and thus the Compton
correction results in  a braking force, since in the scattering of more
energetic
photons on relatively less energetic electrons, a part of energy transferred
to electrons goes to increase the gas thermal energy. Although this
effect tends to lower the radiative force, in a way similar to
the decrease of the scattering cross--section in the Klein--Nishina regime,
it is a completely different phenomenon. The region on the
($M_{env}$, $\mdot$) plane for the existence of stationary, supersonic
solutions permitted by condition (12) lies below the full line in figure 7,
which represents the values of $\mdot_{min}$ obtained equating
expression (12) to zero. This expression was evaluated at the sonic
point for the different numerical models we have computed. The dashed
curve of figure 7 shows the actual values of $\mdot$ for the same solutions.
Although numerical difficulties prevented us to reach the lower possible
value for the mass loss rate where the two curves cross, there is no doubt
that a crossing occurs at $\mdot_{min}\sim 1\div 3$. No stationary, supersonic
winds can exist with $\mdot<\mdot_{min}$. In figure 8
the location of
the sonic radius is plotted vs. $\mdot$: as it should be expected $r_s$
steeply increases for low enough values of $\mdot$.
We note that, in the absence of Comptonization, the sonic radius would
monotonically decrease for decreasing $\mdot$, so the minimum in figure 8
marks the range of $\mdot$ at which Compton heating starts to dominate.

The request that the outflow can be described by our stationary model places
also an upper limit on both $\mdot$ and the total envelope mass, $M_0$,
at the time the wind starts. Keeping in mind that
$M_{env}$ is the mass between the base of the burning shell and the sonic
radius, the variation per unit time of the total envelope mass (which includes
the nuclear processed material laying below the burning shell) is just
$dM_{tot}/dt = -\Mdot$. We assume a simple relation between $M_{env}$ and
$M_{tot}$ of the form $M_{env} = M_{tot}(1-t/t_{nuc})$, which just
states that on a timescale $t_{nuc}$ all the helium will be burned out;
furthermore we approximate the $\log\mdot$--$\log M_{env}$ dependence
with a linear law, $\mdot  = A\, M_{env}^\alpha$. Expressing the
initial differential equation in terms of $M_{tot}$ only, we get the solution
$$M_{tot}=M_0\left\{1-BM_0^\alpha\left[ 1 - \left(1-{t\over{t_{nuc}}}\right)^
{\alpha +1}\right]\right\}^{1/(1-\alpha)}\, .$$
For a model to be stationary the mass lost
in the wind must be much smaller than $M_{tot}$, which implies that
$BM_0^\alpha
\ll 1$. In our case, the limiting initial mass turns out to be $M_0\sim 1.8
\times 10^{26}$ g, corresponding to a a maximum mass loss rate $\mdot\sim
630$, or, $\Mdot\sim 1.2\times 10^{20}$ g/s. All models we computed are
below this critical value of $\mdot$. Moreover, even if the $M_0$
exceeds the above mentioned limit, after an initial,
high mass loss, unstationary phase during which our model cannot be applied,
the wind will enter the parameter range where the outflow can be reliably
treated as a stationary one.

The summary of our series of models is given in table 1, where some global
quantities are presented, namely $\Menv$, $v_{\infty}$, photospheric and
last scattering radii, $R_{ph}$ and $R_{es}$ where $\tau_{eff}=1$ and $\tau =1$
respectively,
matter temperature at $R_{ph}$, $T_{ph}^m$, the characteristic
timescale for expelling the whole envelope, $t_{wind} = \Menv/\Mdot$, and
the nuclear timescale, $t_{nuc}$.
Observations show that the timescale of the expansion phase, over
which the luminosity stays
nearly constant around the maximum (i.e. near to the Eddington value), is
about few seconds. The values of $t_{nuc}$ in Table 1 are indeed of the right
order of magnitude, and it should be also taken into account that $t_{nuc}$
is an upper limit for the duration of the expansion, because not all the
helium may actually be burned out. As it appears from the table, $t_{wind}$
is much longer than $t_{nuc}$, showing that the decrease of the envelope mass
in time is due to
nuclear burning, the mass lost in the wind being less important by far.

In order to access the relevance
of chemical composition on the global properties of our solutions, a series of
nearly pure helium models has been computed, $X=0.05$, $Y=0.90$ and $Z=0.05$.
Results are presented in table 2. In general, for the same $\mdot$,
the envelope tends to be more
compact with respect to the one with solar composition, all relevant
radii are smaller and also $\Menv$ is lower. Variations, however, are within
a factor 2 and timescales are very nearly the same.

Although no frequency dependent calculations are presented here, the
comparison between our results and spectral data of X--ray bursts can
actually yield some useful informations. To illustrate this let us
refer to the EXOSAT observations of 4U/MXB 1820-30 as presented by Haberl
{\it et al.\/} This source is a binary with an 11 minute orbital period,
consistent with a scenario in which the secondary is a low--mass, helium--rich
star (Rappaport {\it et al.\/} 1987). Since we expect the transferred
material to be nearly pure helium, our helium--rich set of models will be used.
Fitting the observed bursts spectra with a planckian law gives
the color temperature $T_{col}$, which can be used to derive an estimate of
the envelope radius $R_{col}$, via the relation $L = 4\pi R_{col}^2\sigma
T_{col}^4$. While the possible anisotropy of the radiation emitted during a
burst, due to the presence of the accretion disk, may be relevant in evaluating
the bolometric luminosity (see e.g. Lapidus \& Sunyaev 1985), it does not
influence spectral data.
The variation of $R_{col}$ with time
is taken as an indication of the envelope expansion and successive contraction
during the burst. It should be noted, however, that, while the previous
argument is correct, $R_{col}$
is not directly related to any physically meaningful radius and
no ``color'' radius can be extracted out of our models. The radius which
does have an evident physical meaning, as far as spectral formation is
concerned, is the photospheric radius $R_{ph}$ since it is here that the
blackbody spectrum originates, with a temperature equal to the matter
temperature $T_{ph}^m$; clearly the emergent spectrum will not be
planckian because of Compton scatterings in the outer, translucent layers.
The $T_{ph}^m$--$R_{ph}$ relation for our helium models is shown in figure 9;
$\Mdot$ decreases moving to higher temperatures along the curve.
As it should be expected, our data deviate from the analytical law
but the trend is the same and, moreover, the observed increase in time
of the color temperature (see e.g. Haberl {\it et al.} figure 4)
corresponds to an increase of $T_{ph}^m$ for
decreasing $\Mdot$ in our data, providing further evidence that the time
evolution
of the wind can be mimicked as a succession of stationary models with
decreasing $\Menv$.

Our results can also be used to construct a $R_{col}$--$T_{col}$ plot,
in the same way as with observational data, and this enables us
to derive an estimate of the spectral hardening, without any need of frequency
dependent calculations. In fact, by introducing
a hardening factor $\gamma = T_{col}/T_{ph}^m$, we have
$$4\pi R_{col}^2\sigma (\gamma T_{ph}^m)^4 = L\, ,\eqno(13)$$
where $L$ can be safely assumed to be the Eddington luminosity.
The value of $\gamma$ can be derived asking that
$$\gamma T_{ph}^m\left\vert_{\mdot_{min}}\right. = T_{col}\left\vert_{max}
\right.$$
where $T_{col}\vert_{max}$ is the maximum observed color temperature, which
is $\simeq 3$ keV for the bursts in 4U/MXB 1820-30 analyzed by Haberl {\it
et al.\/} We emphasize that the choice of the last point as the fiducial one
is based on the existence of a minimum value of $\Mdot$ which is assumed
to be reached at the end of the expanding envelope phase.
The hardening factor obtained in such a way turns out to be
$\gamma\sim 1.53$ and, since $\gamma > 1$, we expect
a genuine hardening of the spectrum as radiation propagates through the
extended spherical shell $R_{ph}<R<R_{es}$. This result does not contradict the
previous finding of Lapidus who obtained a softening (rather than hardening)
factor $\sim 0.25 \div 0.7$,
solving the frequency dependent transfer problem on a fixed hydrodynamical
background with a solar chemical composition. This is because, in his
investigation, the softening factor
was defined as $\gamma_{soft}=T_{col}/T_{eff}$, $T_{eff}$ being the effective
temperature
at the neutron star surface, $T_{eff}\sim 2 \ {\rm keV}\magcir
T_{ph}^m\sim 0.3\div 2$ keV. The scaling between the two factors is
just $T_{eff}/T_{ph}^m\sim 1\div 7$, in agreement with
$\gamma/\gamma_{soft}\sim 2\div 6$. The ratio $T_{col}/T_{eff}$
is, typically, $\sim 1.5$ in a static atmosphere (see e.g. London {\it
et al.\/}) and then abruptly drops to $\mincir 0.7$ at the onset of the wind
phase, but in both situations the blackbody spectrum, produced at the
thermalization radius, will be then hardened by Comptonization.
In figure 10 the curve derived from equation (13), with $\gamma = 1.53$,
is superimposed to the
data of branch {\it b\/} of Haberl {\it et al.\/} figure 4, here
represented by the shaded area. We restrict our attention to branch {\it b}
because it can be taken as representative of the quasi--stationary wind
phase which is modeled by our solutions. Within this framework,
it is natural to interpret the intersection between our curve and
the left border of the shaded area as the point in the
parameter space where the quasi--stationary wind phase
begins. The intersection
point corresponds to a model with $\mdot\sim 90$, $\Menv\sim
9\times 10^{23}$ g and the latter value may be assumed as the total
mass of the envelope at the onset of the burst, $M_0$.
As can be seen from table 1,
the final envelope mass, corresponding to the minimum possible $\mdot$, is
$\sim 2\times 10^{21} \ {\rm g}\ll M_0$, so that nearly all the
helium
is burned out. The time required for this process is $\sim t_{nuc}\sim 10$ s
which is close to the observed duration, $\sim 5$ s, of the
quasi--stationary phase.
Taking the interburst time $\Delta t = 1.1\times
10^4$ s, we get an estimate of the neutron star accretion rate: $\Mdot_{acc}
\sim 10^{-6} \ M_\odot/yr$.
The present estimate for $\Mdot_{acc}$ turns out to be quite
high in comparison with the values discussed in connection with model
neutron star atmospheres with nuclear burnings (see e.g. Ayasli \& Joss 1982;
Fushiki \& Lamb 1987). A lower value for $\Mdot_{acc}$ can be obtained relaxing
the hypothesis that $\gamma$ is same for all the wind models. Of course there
is
no physical reason to expect this to be true. It is reasonable, in fact, to
assume that the hardening ratio increases with decreasing envelope mass since
low $\Mdot$ models have a more extended, hotter scattering region
(see e.g. figure 2) where Comptonization is more effective. In this framework
the value of $\gamma$ we have computed should be the maximum one and a lower
limit for the initial envelope mass can be obtained assuming the initial
model to have $\gamma = 1$, in which case $T_{ph}^m = T_{col}\sim 0.5$ keV.
Data in table 2 show that now $M_0\sim 5\times 10^{23}$ g which gives
$\Mdot_{acc}\sim 5\times 10^{-7}M_\odot/yr$, about half our previous estimate.
Still lower accretion rates could be obtained if $\gamma <1$ at the beginning
of the quasi--stationary contraction phase, although a softening of the
spectrum with respect to the blackbody at $T_{ph}^m$ seems unplausible. We
note that the actual dependence of $\gamma$ on $\Menv$ does not affect
the previous conclusions which rely only on the given initial value of the
hardening factor. More severe uncertainties in the determination of
$\Mdot_{acc}$ stem from observational data. In the case of 4U 1820-30, Haberl
{\it et al.\/} reported that, during the first 3 seconds of each burst, the
blackbody fitting to the observed spectrum was rather poor. Since the
contraction phase begins just after $\sim 1$ s, we expect the largest
errors to affect the minimum color temperature which is the key parameter
in selecting the starting wind model. From the discussion in Haberl {\it et
al.\/} about the fitting of early spectra, we surmise
that a value of $T_{col}$ higher than 0.5 keV could
be more appropriate. These authors state that, at the onset of the contraction
phase, spectra show a broad maximum and were
well fitted by a power law plus a bremsstrahlung. Such spectra are typically
produced in rather dense, expanded envelopes with little or no Comptonization
and a substantial free--free emission outside the thermalization radius, like
what is expected in our more massive wind models. If $T_{col}$ has to serve
as a measure of the photospheric temperature, as in our case, a fit of the
exponential tail would
be more significant, since it is this part of the spectrum which originates
at the
thermalization radius, and will give a larger $T_{col}$. The resulting
accretion rate can be much reduced because $M_{env}$ decreases rather steeply
with $1/T_{ph}^m = \gamma/T_{col}$. For instance,
assuming $T_{col}^{min}\sim 1$
keV and $\gamma = 1$, we get $\Mdot_{acc}\sim 10^{-7}M_\odot/yr$, which is
still rather high. On the other hand, an application of the same technique
to all other burst sources with photospheric expansion (Lapidus, Nobili \&
Turolla 1994) produced much lower values of the accretion rate, $\Mdot_{acc}
\sim 10^{-8}\div 10^{-9}M_\odot/yr$, supporting the current idea that 4U
1820-30 is a peculiar object.

Although for all present estimates $\Mdot_{acc}$ is highly supercritical,
no significant release of gravitational
luminosity, $L_{acc}\simeq GM_*\Mdot_{acc}/c^2 R_*$, occurs in the interburst
phase because the gas has no time to cool. The inner part of the flow, in fact,
is optically thick and the appropriate cooling time is the adiabatic time
(see e.g. Bildsten 1993)
$$t_{cool}^{ad}\sim {{c_p\kes (\rhoz r)^2}\over{caT^3}}\sim
10^{11} \left({{\rhoz}\over{10^6 \ {\rm g}}}\right)^2
\left({{T}\over{10^8 \ {\rm K}}}\right)^{-3} \ {\rm s}\, ,$$
where $c_p$ is the specific heat at constant pressure. Since
$\rhoz T^{-3/2}$ is nearly constant,
the cooling time can be computed using the values of $\rhoz$ and $T$ which
correspond to the helium ignition at the base of the accretion flow.
For $\mdot\sim 90$, it is
$\rhoz\sim 10^6 \ {\rm g\, cm}^{-3}$, $T\sim 6\times 10^9$ K, and we get
$$t_{cool}^{ad}\sim 5\times 10^5\ {\rm s}\gg t_{acc}\sim 10^4 \ {\rm s}\, .$$
It follows, then, that the heat produced by the conversion of gravitational
potential energy can not be radiated away in a time $t_{acc}$ and must go
to increase the gas internal energy. This in turn implies that the accretion
process can not be regarded as stationary.
Only a small fraction of $L_{acc}$ is
expected to escape to infinity while the progressive heating of the inner
gas layers produces, at the end, the helium flash. It is either this fraction
of $L_{acc}$ or the stationary
hydrogen burning on the surface of the neutron star (see Ayasli \& Joss,
Taam {\it et al.\/}) which are responsible for the observed persistent
X--ray luminosity, $\sim 0.1 L_E$. The fact that the persistent luminosity
has been observed to be higher ($\sim L_E$) when the source was burst
inactive (see e.g. Stella {\it et al.\/} 1984) could be explained in terms
of a lower accretion rate, for which the process is stationary. Under such
conditions the flow has time to radiate away all the gravitational energy
and a larger luminosity can be produced with a smaller $\Mdot_{acc}$.

A simple argument can be
used to estimate the initial temperature at the base of the envelope needed
to ignite the helium after a given amount
of material is accreted. If we assume
that the inner accretion layers are in hydrostatic equilibrium and the gas
is heated adiabatically, it can be shown that $\Menv\propto T^{5/2}$.
At the beginning of the accretion process the envelope
mass is $\sim 2\times 10^{21}$ g, and $T$ should be $\sim  10^9$ K,
in order
to reach $\sim 6\times 10^9$ K when $\Menv\sim 9\times 10^{23}$ g. This implies
that the deeper layers should cool from $\sim 2\times 10^9$ K, which is the
value of $T_b$ when the expanded phase ends, to $ 10^9$ K in the
burst decay time.

We point out that the comparison of our solutions with the data of
4U/MXB1820-30 was primarily intended as a test on the viability of our
model and no attempt was made here to really fit the
observational data by varying the free parameters of the model, i.e.
the neutron star mass, radius and the chemical composition of the
outflowing gas.

\beginsection IV. CONCLUSIONS

We have presented a new model for stationary winds from neutron stars which
accounts properly for the relevant radiative processes, and handles
correctly
the radiative transfer in all regimes, from diffusion to radial streaming.
Unlike previous investigations on this subject, the energy input rate
is not treated as a free parameter, but is consistently derived from
thermonuclear helium burning at the base of the envelope. In accordance
with generally accepted scenarios, the energy released in excess of the
Eddington luminosity is converted into the kinetic energy of the outflowing
envelope, so that the radiative flux seen by a distant observer is always
very close to the Eddington value. At the present
stage only the frequency--integrated problem was solved.
Frequency--dependent calculations are in progress and will be published later.
We have found that each model is characterized by only one free parameter:
either the mass loss rate $\Mdot$, or the total envelope
mass $M_{env}$. It is shown that, due to the effects of Comptonization,
there exists a lower limit for $\Mdot$, i.e. stationary winds can exist
only with
$\Mdot$ larger than $\Mdot_{min} \, \approx \, \Mdot_{E}$.  We discussed
how the sequence of our models may be used to mimic the temporal evolution
of a strong X--ray burst during the expansion/contraction phase.
Matching of our models with observational data of
4U/MXB 1820-30 results in a spectral hardening factor
$\gamma\sim  1.5$, in accordance with the theoretical prediction that
a genuine hardening of the spectrum occurs as radiation propagates from the
photosphere outwards. We were able, also, to get an estimate of the accretion
rate from the companion star between two successive strong type I bursts in
this source, $\Mdot_{acc}\sim 10^{-7}\div 10^{-6} \ M_\odot$/yr.

Further work should be aimed to compute of a whole grid of wind
models, varying $M_*$, $R_*$, chemical composition, and to include a more
complete treatment of nuclear reactions. In any case, in order to follow
the burst evolution outside the quasi--stationary phase a full, time
dependent approach is needed.

\beginsection ACKNOWLEDGMENTS

We thanks A.~Fabian, M.~Rees, P.~Podsiadlowski for useful discussions and
an anonymous referee for some helpful comments.
One of us (I.~L.) gratefully acknowledges financial support from
Consiglio Nazionale delle Ricerche (Gruppo Nazionale di Astronomia) and
Royal Astronomical Society. He is also indebted to the Department of Physics,
University of Padova, for kind hospitality during his stay.

\vfill\eject

\beginsection REFERENCES

\ref{Ayasli, S., \& Joss, P.C., 1982, ApJ, 256, 637}
\ref{Bildsten, L. 1993, ApJ, to appear}
\ref{Cassinelli, J.P. 1979, ARA\&A, 17, 275}
\ref{Chandrasekhar, S. 1939, Stellar Structure, (New York: Dover)}
\ref{Cheng, A.Y.S., \& O'Dell, S.L. 1981, ApJ, 251, L49}
\ref{Clayton, D.D. 1968, Principles of Stellar Evolution and
Nucleosynthesis, (New York: Mc Graw--Hill)}
\ref{Duncan, R.C., Shapiro, S.L., \& Wasserman, I. 1986, ApJ, 309, 141}
\ref{Ebisuzaki, T., Hanawa, T., \& Sugimoto, D. 1983, PASJ, 35, 17}
\ref{Flammang, R.A. 1982, MNRAS, 199, 833}
\ref{Fushiki, I., \& Lamb, D. Q. 1987, ApJ, 323, L55}
\ref{Goodman, J. 1986, ApJ, 308, L47}
\ref{Haberl, F., Stella, L., White, N.E., Priedhorsky, W.C., \& Gottwald, M.
1987, ApJ, 314, 266}
\ref{Joss, P.C., \& Melia, F. 1987, ApJ, 312, 700}
\ref{Kato, M. 1983, PASJ, 35, 33}
\ref{Lapidus, I.I., \& Sunyaev, R.A. 1985, MNRAS, 217, 291}
\ref{Lapidus, I.I., Sunyaev, R.A., \& Titarchuk, L.G. 1986,
Pis'ma Astron. Zh. 12, 918 (Sov. Astron. Lett., 12, 383 (1987))}
\ref{Lapidus, I.I. 1991, ApJ, 377, L93}
\ref{Lapidus, I., Nobili, L., \& Turolla, R. 1994, ApJ submitted}
\ref{Lewin, W.H.G., Vacca, W.D., \& Basinska, E. 1984, ApJ,
277, L57}
\ref{Lewin, W.H.G., Van Paradijs, J., \& Taam, R.E. 1993, Space Sci.
Rev., 62, 223}
\ref{London, R.A., Taam, R.E., \& Howard, W.E. 1986, ApJ, 306, 170}
\ref{Meier, D.L. 1982a, ApJ, 256, 681}
\ref{Meier, D.L. 1982b, ApJ, 256, 693}
\ref{Meier, D.L. 1982c, ApJ, 256, 706}
\ref{Nobili, L., \& Turolla, R. 1988, ApJ, 333, 248}
\ref{Nobili, L., Turolla, R., \& Zampieri, L. 1991, ApJ, 383, 250}
\ref{Nobili, L., Turolla, R., \& Zampieri, L. 1993, ApJ, 404, 686}
\ref{O'Dell, S.L. 1981, ApJ, 243, L147}
\ref{Paczy\'nski, B. 1983, ApJ, 267, 315}
\ref{Paczy\'nski, B. 1986, ApJ, 308, L43}
\ref{Paczy\'nski, B., \& Anderson, N. 1986, ApJ, 302, 1}
\ref{Paczy\'nski, B., \& Pr\'oszy\'nski, M. 1986, ApJ, 302, 519}
\ref{Paczy\'nski, B. 1990, ApJ, 363, 218}
\ref{Park, M.--G., \& Ostriker, J.P. 1989, ApJ, 347, 679}
\ref{Payne, D.G, \& Blandford, R.D. 1981, MNRAS, 196, 781}
\ref{Quinn, T., \&  Paczy\'nski, B. 1985, ApJ, 289, 634}
\ref{Rappaport, S., Nelson, L., Joss, P., \& Ma, C.--P. 1987, ApJ, 322, 842}
\ref{Salpeter, E.E., \&  Shapiro, S.L. 1981, ApJ, 251. 311}
\ref{Shestakov, A.I., Kershaw, D.S. \& Prasad, M.K. 1988, J. Quant.
Spectrosc. Radiat. Transfer, 40, 577}
\ref{Stella, L., Kahn, S.M., \& Grindlay, J.E. 1984, ApJ, 282, 713}
\ref{Taam, R.E. 1985, Ann. Rev. Nucl. Particle Sci.,  35, 1}
\ref{Taam, R.E., Woosley, S.E., Weaver, T.A., \& Lamb, D.Q. 1993,
ApJ, 413, 324}
\ref{Tawara, Y., {\it et al.\/} 1984, ApJ, 276, L41}
\ref{Tawara, Y., Hayakawa, S., \& Kii, T., 1984, PASJ, 36, 845}
\ref{Thorne, K.S. 1981, MNRAS, 194, 439}
\ref{Titarchuk, L. 1993, ApJ, submitted}
\ref{Turolla, R., Nobili, L., \& Calvani, M. 1986, ApJ, 303, 573}
\ref{Turolla, R., \& Nobili, L. 1988, MNRAS, 303, 573}
\ref{Zampieri, L., Turolla, R., \& Treves, A. 1993, ApJ, 419, 311}

\vfill\eject

\beginsection FIGURE CAPTIONS

\refig{Figure 1.\quad Bulk velocity (continuous line), sound speed
(dashed line) and density (in g/cm$^3$, dotted line) versus radius
for the models with $\mdot = 2.8$ (a), $\mdot = 50.2$ (b) and $\mdot = 102.3$
(c); both scales are logarithmic and a filled dot marks the
fitting radius.}
\medskip
\refig{Figure 2.\quad Same as figure 1 for matter (continuous line) and
radiation (dashed line) temperatures.}
\medskip
\refig{Figure 3.\quad  Luminosity, as measured by the comoving observer
in Eddington units, versus radius for $\mdot = 102.3$ (continuous line),
$\mdot = 50.2$ (dashed line) and $\mdot = 2.8$ (dotted line).}
\medskip
\refig{Figure 4.\quad Electron scattering (continuous line)
and effective (dashed line) optical depths for the model with $\mdot = 50.2$;
the run of $\tau v$ (dotted line) is also shown.}
\medskip
\refig{Figure 5.\quad The run of radiation moments, $w_0$ (upper curve)
and $w_1$ (lower curve), for the model with $\mdot = 50.2$.}
\medskip
\refig{Figure 6.\quad The variation of $\Menv$ versus $\mdot$.}
\medskip
\refig{Figure 7.\quad The $\mdot$--$\Menv$ relation for a sample of models
(continuous line) together with the limit given by equation (12) (dashed
line). Only the low $\mdot$ range is shown; see text for details.}
\medskip
\refig{Figure 8.\quad The variation of sonic radius, $R_s$, versus
$\mdot$.}
\medskip
\refig{Figure 9.\quad The $T_{ph}^m$--$R_{ph}$ relation for helium solutions;
the photosphere is defined by  the condition $\tau_{eff}= 1$.}
\medskip
\refig{Figure 10.\quad
The comparison between the relation $4\pi R_{col}^2\sigma (\gamma T_{ph}^m)^4=
L$, for helium models with $\gamma = 1.53$, and the data of branch {\it b}
of Haberl {\it et al.\/} for 4U/MXB 1820-30 (shaded area).}

\vfill\eject
%
\baselineskip=16truept
\centerline{Table 1}\medskip
\centerline{Characteristic Parameters for Selected Solar Composition
Models}\bigskip
$$\vbox{\tabskip=1em plus2em minus.5em
\halign to\hsize{#\hfil &\hfil # \hfil & \hfil # \hfil & \hfil # \hfil &
 \hfil # \hfil  & \hfil # \hfil & \hfil # \hfil &
\hfil # \hfil & \hfil # \hfil &\hfil # \hfil &\hfil # \hfil\cr
& $\Mdot $ & $\Menv $ & $v_\infty$
& $R_{ph}$ & $R_{es}$
& $T_{ph}^m$ & $t_{wind}$ & $t_{nuc}$ & & \cr
& $(\Mdot_{E})$  & $(10^{22}\, {\rm g}) $ & $(10^{-3}c)$
& $(10^3\, {\rm km})$ & $(10^3\, {\rm km})$
& $({\rm keV})$ &({\rm s}) & ({\rm s}) & & \cr
%
%
\noalign{\medskip}
 &  139.2 &  195.8 &    1.20 &   19.37 &  180.71 &    0.06 &3157 &  10& &\cr
 &  124.7 &  174.4 &    1.23 &   14.82 &  158.69 &    0.08 &2937 &  10& &\cr
 &  113.0 &  156.7 &    1.25 &   11.74 &  142.40 &    0.09 &2668 &  10& &\cr
 &  102.3 &  140.3 &    1.28 &    9.27 &  127.60 &    0.11 &2399 &  10& &\cr
 &   89.8 &  121.3 &    1.34 &    6.67 &  107.63 &    0.13 &2156 &  10& &\cr
 &   79.1 &  104.1 &    1.48 &    4.63 &   86.77 &    0.17 &1920 &  10& &\cr
 &   68.7 &   87.3 &    1.66 &    3.33 &   67.61 &    0.20 &1688 &  9 & &\cr
 &   59.3 &   72.0 &    1.81 &    2.55 &   53.75 &    0.24 &1460 &  9& &\cr
 &   53.9 &   63.5 &    1.92 &    2.27 &   46.53 &    0.26 &1345 &   9& &\cr
 &   50.2 &   57.1 &    2.00 &    2.03 &   41.97 &    0.27 &1242 &  8& &\cr
 &   40.2 &   40.9 &    2.21 &    1.51 &   31.12 &    0.33 & 996 &   7& &\cr
 &   34.0 &   30.9 &    2.32 &    1.22 &   25.74 &    0.37 & 847 &  7& &\cr
 &   30.4 &   25.6 &    2.42 &    1.04 &   22.49 &    0.41 & 754 &  6& &\cr
 &   26.5 &   20.0 &    2.51 &    0.88 &   19.40 &    0.45 & 672 &  5& &\cr
 &   19.9 &   11.5 &    2.81 &    0.59 &   13.80 &    0.57 & 519 &   4& &\cr
 &   16.3 &    7.8 &    2.94 &    0.46 &   11.36 &    0.66 & 455 &  3& &\cr
 &   14.3 &    6.0 &    3.04 &    0.39 &   10.00 &    0.73 & 419 &   3& &\cr
 &   11.3 &    3.9 &    3.16 &    0.31 &    8.17 &    0.85 & 385 &   2& &\cr
 &    8.8 &    2.5 &    3.17 &    0.26 &    6.94 &    0.95 & 370 &   2& &\cr
 &    6.7 &    1.7 &    3.32 &    0.20 &    5.62 &    1.09 & 364 &   2& &\cr
 &    5.9 &    1.4 &    3.32 &    0.19 &    5.26 &    1.14 & 369 &   2& &\cr
 &    4.8 &    1.1 &    3.28 &    0.18 &    4.77 &    1.18 & 396 &   1& &\cr
 &    2.8 &    0.7 &    3.33 &    0.15 &    3.69 &    1.31 & 515 &   1& &\cr
 }}$$

\vfill\eject
%
\centerline{Table 2}\medskip
\centerline{Characteristic Parameters for Selected Helium Models}\bigskip
$$\vbox{\tabskip=1em plus2em minus.5em
\halign to\hsize{#\hfil &\hfil # \hfil & \hfil # \hfil & \hfil # \hfil &
 \hfil # \hfil  & \hfil # \hfil & \hfil # \hfil &
\hfil # \hfil & \hfil # \hfil &\hfil # \hfil &\hfil # \hfil\cr
& $\dot M $ & $M_{env} $ & $v_\infty$
& $R_{ph}$ & $R_{es}$
& $T_{ph}^m$ & $t_{wind}$ & $t_{nuc}$ & & \cr
& $(\dot M_{E})$  & $(10^{22}\, {\rm g}) $ & $(10^{-3}c)$
& $(10^3\, {\rm km})$ & $(10^3\, {\rm km})$
& $({\rm keV})$ & ({\rm s}) & ({\rm s}) & & \cr
%
%
\noalign{\medskip}
 &  130.3 &  137.3 &    1.51 &    3.55 &   47.92 &    0.23 &2812 &  13& &\cr
 &  123.5 &  129.7 &    1.55 &    3.24 &   44.60 &    0.25 &2692 &  13& &\cr
 &  116.3 &  121.5 &    1.56 &    3.00 &   42.15 &    0.26 &2592 &  13& &\cr
 &  103.3 &  106.3 &    1.68 &    2.47 &   34.81 &    0.29 &2361 &  13& &\cr
 &   92.9 &   94.1 &    1.77 &    2.06 &   30.05 &    0.33 &2110 &  13& &\cr
 &   82.2 &   81.1 &    1.87 &    1.73 &   25.49 &    0.37 &1874 &  12& &\cr
 &   70.1 &   66.6 &    1.97 &    1.43 &   21.04 &    0.41 &1656 &  12& &\cr
 &   62.1 &   56.5 &    2.14 &    1.17 &   17.24 &    0.47 &1424 &  11& &\cr
 &   56.0 &   48.9 &    2.16 &    1.06 &   15.68 &    0.50 &1324 &  11& &\cr
 &   51.5 &   43.2 &    2.19 &    0.96 &   14.42 &    0.53 &1220 &  10& &\cr
 &   40.5 &   29.5 &    2.42 &    0.70 &   10.54 &    0.64 & 961 &  9& &\cr
 &   34.8 &   22.5 &    2.50 &    0.57 &    8.99 &    0.71 & 835 &  8& &\cr
 &   30.9 &   17.9 &    2.57 &    0.49 &    7.91 &    0.78 & 750 &  7& &\cr
 &   24.5 &   11.0 &    2.88 &    0.35 &    5.78 &    0.97 & 579 &  5& &\cr
 &   20.3 &    7.1 &    2.91 &    0.28 &    4.94 &    1.09 & 513 &  4& &\cr
 &   16.2 &    4.0 &    2.90 &    0.22 &    4.19 &    1.24 & 455 &  3& &\cr
 &   12.4 &    2.0 &    2.88 &    0.17 &    3.50 &    1.46 & 405 &   2& &\cr
 &   10.4 &    1.3 &    2.82 &    0.15 &    3.20 &    1.57 & 399 &   1& &\cr
 &    7.1 &    0.6 &    2.74 &    0.12 &    2.70 &    1.81 & 395 &   1& &\cr
 &    6.6 &    0.5 &    2.73 &    0.11 &    2.61 &    1.86 & 396 &   1& &\cr
 &    5.9 &    0.4 &    2.68 &    0.11 &    2.54 &    1.89 & 412 &   1& &\cr
 &    4.8 &    0.3 &    2.60 &    0.10 &    2.42 &    1.97 & 442 &   1& &\cr
 &    4.2 &    0.2 &    2.56 &    0.10 &    2.31 &    2.03 & 467 &   1& &\cr
 }}$$

\vfill\eject

\bye